\documentclass{ws-procs975x65}
\newcommand{\nup}{$\nu_{\rm peak}$}
\newcommand{\lsim}{{\lower.5ex\hbox{$\; \buildrel < \over \sim \;$}}}
\newcommand{\gsim}{{\lower.5ex\hbox{$\; \buildrel > \over \sim \;$}}}

\begin{document}

\title{A SIMPLIFIED VIEW OF BLAZARS: \\ COMPARISON WITH MULTI-FREQUENCY OBSERVATIONS}

\author{P. GIOMMI$^*$}

\address{ASI Science Data Center, ASDC \\
Italian Space Agency, Via del Politecnico snc 
I-00133 Roma, Italy\\
$^*$E-mail: paolo.giommi@asdc.asi.it}

\author{P. PADOVANI}
 
\address{European Southern Observatory, Karl-Schwarzschild-Str. 2,\\
D-85748 Garching bei M\"unchen, Germany\\
}

\begin{abstract}
We have recently proposed a new scenario where blazars are classified as
flat-spectrum radio quasars or BL Lacs according to the prescriptions of unified schemes, 
and to a varying combination of Doppler boosted radiation from the jet, emission from the
accretion disk, the broad line region, and light from the host galaxy. This mix of different
components leads to strong selection effects, which are 
properly taken into account in our scheme. We describe here the main features of our approach, 
which solves many long-standing issues of blazar research, give the most important results, 
and discuss its implications and testable predictions. 
\end{abstract}


\bodymatter
\section{The problem}

Blazars are radio loud active galactic nuclei (AGN) pointing their jets in the direction of the observer 
\cite{bla78,UP95}. Historically, they have been divided in two main subclasses, whose 
major difference is in their optical properties: 1) Flat Spectrum Radio Quasars (FSRQs), which 
show strong, broad emission lines in their optical spectra, just like classical quasars; and 2) 
BL Lacertae objects (BL Lacs), which at most show weak emission
lines, sometimes display absorption features, and in some cases can be completely featureless. 
These two classes display many other differences in their: 1) extended radio powers; 2) redshift distributions; 3) cosmological evolutions; 4) 
synchrotron peak energy (\nup) distributions. Finally, radio, X-ray, and $\gamma$-ray 
selected  blazar samples show a very different mix of FSRQs and BL Lacs. 

Some of these differences have been explained by the so-called unified schemes, which posit 
that BL Lacs and FSRQs are simply Fanaroff-Riley (FR) I/low excitation radio galaxies (LERGs) and 
FR II/high excitation radio galaxies (HERGs) with their jets forming a small angle 
with respect to the line of sight \cite{UP95}. However, unified schemes per se cannot account 
for transitional objects (that is, sources which sometimes move from one class to the other), 
which include, for example, even BL Lacertae itself and 3C 279, the different \nup\ distributions 
of FSRQs and BL Lacs (with the latter reaching much higher values), and the very different 
evolutions of radio and X-ray selected BL Lacs. 

\section{Our solution}

To explain all of the above differences, 
in a series of papers \cite{SimplifiedI,SimplifiedII,SimplifiedIII} we have proposed a new 
scenario (dubbed a {\textit {simplified blazar view}), which we kept as simple as possible 
and tied as much as possible to observational data. Our starting point is the 
observational fact that the optical spectrum of blazars is the sum of three
components: a non-thermal, jet related one, one due to the
accretion disk, and one host-galaxy related. Different mixes of these
components determine the appearance of the optical spectrum and therefore
the classification of sources in FSRQs (dominated by strong lines), BL Lacs
(with diluted, weak lines, even if a standard accretion disk is present), and
radio-galaxies (where the host galaxy swamps both the thermal and
non-thermal nuclear emission). Our novel approach assumes a unique
non-thermal ``engine'' for the two classes (based on a simple homogeneous synchrotron 
model), whereas the disk can be different. Namely, we 
associate a standard accretion disk only with HERGs, given that LERGs either do not possess one, 
or if they do, it is much less efficient (i.e., of the
Advection Dominated Accretion Flow [ADAF] type) \cite{ev06}. The other novel component
is a single luminosity function (LF) whose evolution depends on radio power, with lower luminosity 
($P_{\rm r} \lsim 10^{26}$ W/Hz) radio sources displaying a much weaker cosmological evolution 
than high luminosity ones. Further ingredients include the intrinsic (before dilution from the jet)
distributions of the equivalent width (EW) of the broad lines, a ``standard'' elliptical host galaxy, 
and a disk-to-jet power ratio distribution, all derived from observations \cite{SimplifiedI}. 

\begin{figure}[b]%
\begin{center}
  \parbox{2.4in}{\epsfig{figure=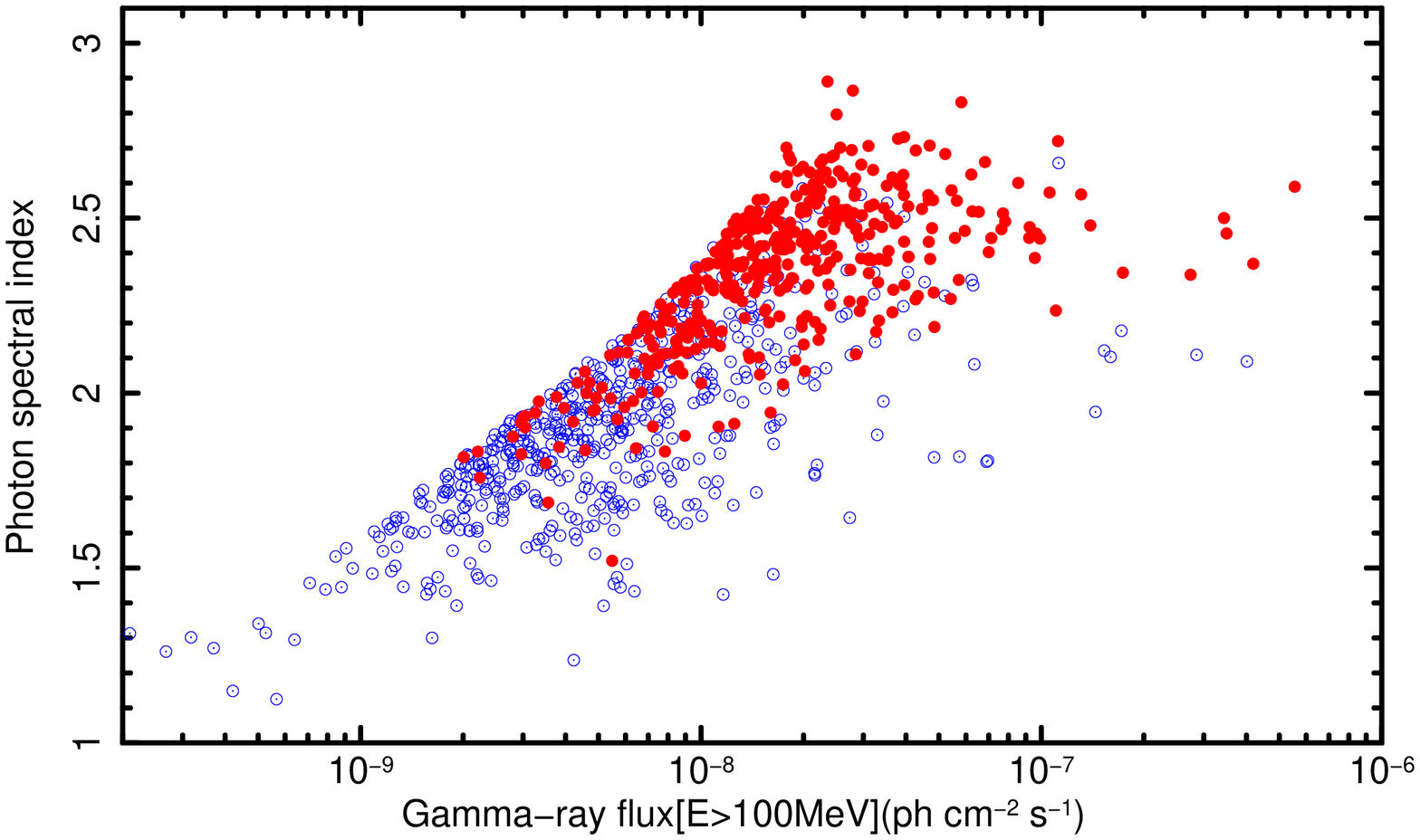,angle=-90,width=2.6in}}
  \hspace*{12pt}
  \parbox{2.3in}{\epsfig{figure=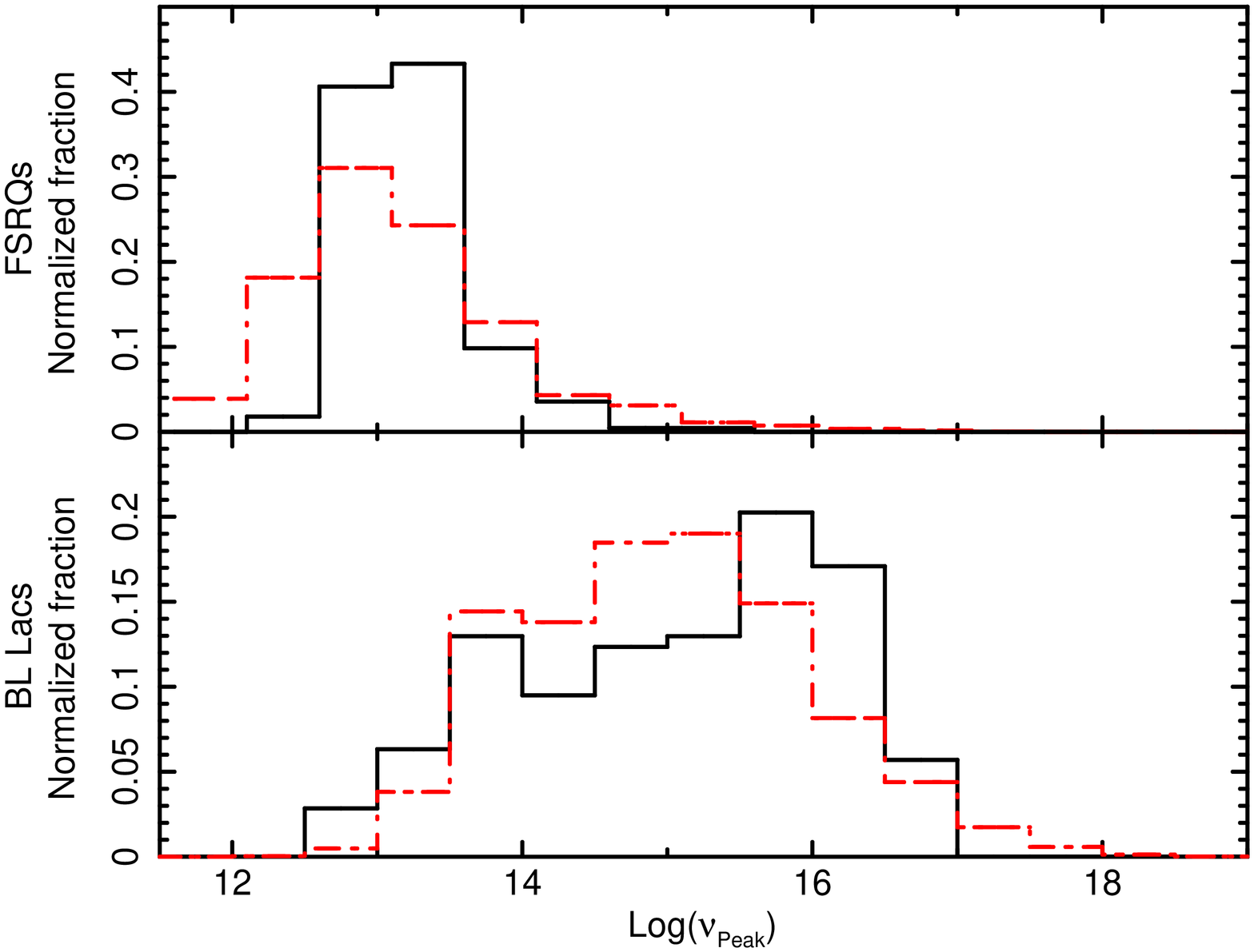,angle=-90,width=2.3in}}
  \caption{{\it Left panel:} the spectral index of blazars in our simulated survey of 1,000 $\gamma$-ray sources vs. the $\gamma$-ray flux. FSRQs are plotted as filled red circles, BL Lacs as open 
  blue circles. 
   {\it Right panel:}  the distribution of \nup\ for blazars in the {\it Fermi} 2LAC catalogue (black solid 
   histogram) and our simulations (red
  dot-dashed histogram) for FSRQs (top panel) and BL Lacs (bottom panel).}
  \label{fig1}
\end{center}
\end{figure}
We thoroughly tested this new approach using Monte Carlo simulations and showed that it
is consistent with the results of radio and X-ray blazar surveys and provides simple answers 
to most, if not all, long-standing, open issues mentioned above. A key result
is that selection effects play a very important role in the diversity observed in radio and X-ray 
samples (and also in the correlation between luminosity and \nup, the so-called ``blazar 
sequence'')\cite{SimplifiedI,SimplifiedII}. We also extended our approach to the $\gamma$-ray band by 
deriving the $\gamma$-ray to radio flux density ratio, $f_{\gamma}/f_{\rm r}$, from the Compton 
dominance (the ratio between inverse Compton and synchrotron peak
luminosities) and \nup\ for radio selected blazars \cite{SimplifiedIII}. Our $\gamma$-ray 
simulations are in agreement with the 
{\it Fermi}-Large Area Telescope (LAT) survey data, including the observed 
percentages of BL Lacs and FSRQs, the fraction of redshift-less objects and the redshift, \nup, 
and $\gamma$-ray spectral index distributions (see, e.g., Fig. \ref{fig1}). 
\section{Implications}
 \begin{table}
 \caption{The  {\textit {simplified blazar view} scenario.  See  text for details.}}
 \begin{tabular}{lccl}
       & LERG & HERG & viewing angle \\
  \hline
 strong optical jet dilution  &  BL Lac  & {\it BL Lac}$^{(1)}$  & $\theta < \theta_{\rm blazar}$  \\
 weak optical jet dilution & {\it radio galaxy}$^{(2)}$ & FSRQ &  $\theta < \theta_{\rm blazar}$ \\
 misdirected jet   &  radio galaxy &  radio galaxy &  $\theta > \theta_{\rm blazar}$ \\
 \hline
 \multicolumn{4}{l}{\footnotesize {\it Italics} denote ``masquerading" sources (see text), $^{(1)}$misclassified FSRQ,  $^{(2)}$misclassified BL Lac}
  \end{tabular}
 \label{tab:scenario}
 \end{table}
Table \ref{tab:scenario} summarizes the main implications of the {\textit {simplified blazar view}. 
Sources with their jets at small angles w.r.t. the line
of sight ($\theta < \theta_{\rm blazar} \sim 15 -
20^{\circ}$)\cite{UP95}, and therefore dominated by non-thermal emission,
are characterized by low values of the EWs and/or the so-called
Ca H\&K break. However, only LERGs belonging to this class are real
BL Lacs, that is have intrinsically weak emission lines. HERGs, which have an accretion disk 
and therefore display strong emission lines, {\it appear} to show weak lines when these are
swamped by non-thermal emission and are therefore ``masquerading" BL Lacs 
({\it  italics}  in the table), being in reality misclassified FSRQs. Sources with somewhat weaker 
optical jet 
dilution still oriented at small angles show some optical features, like the emission lines
typical of FSRQs. In this category we find also ``masquerading" radio galaxies,
that is sources, which are ``bona fide" blazars but have their non-thermal emission in the 
optical/UV part of the spectrum swamped by the host galaxy. Finally, misdirected jets characterize 
the true radio galaxy population. In short, starting from truly different objects like LERGs and 
HERGs, the combination of viewing angles and strong/weak optical/UV light dilution from the jet 
determines the commonly adopted classification in FSRQs, BL Lacs, or radio galaxies. This purely 
observational approach assigns in a number of cases intrinsically different 
objects to the same class, making the task of understanding blazars more complicated.
It  then turns out that sources so far classified as BL Lacs on the
basis of their observed weak, or undetectable, emission lines are of two
physically different classes: intrinsically weak-lined objects, more common
in X-ray selected samples, and heavily diluted broad-lined sources, more
frequent in radio and $\gamma$-ray selected samples, which explains some of the confusion in
the literature.
\begin{figure}[h]%
\begin{center}
  \parbox{2.1in}{\epsfig{figure=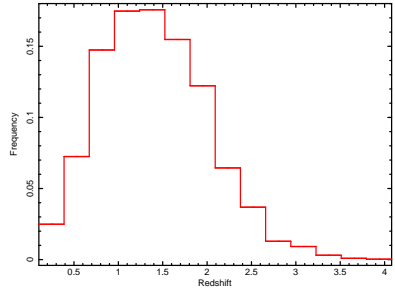,angle=-90,width=2.in}} 
  \caption{The intrinsic redshift distribution of $\gamma$-ray selected featureless BL Lacs.}
  \label{fig2}
\end{center}
\end{figure}
The implications of our results are far reaching. For example, the large fraction of redshift-less 
{\it Fermi}-LAT blazars, which is matched very well by our simulations, must have higher 
redshifts than those of BL Lacs with redshift information, as shown in Fig. \ref{fig2}.
Therefore, one cannot assume for them a value typical of 
the BL Lacs with measured redshift ($z \sim 0.4$) but larger values ($z \approx1-1.5$) need to 
be used. Moreover, most
of these sources are predicted to be quasars with their emission lines
heavily diluted by the non-thermal continuum. These should therefore be
included with the FSRQs when studying number counts, cosmological
evolution, and LFs, since their exclusion will bias the results. Furthermore, 
 ``masquerading" radio galaxies, that is ``bona fide'' blazars having
their optical non-thermal emission swamped by the host galaxy, could be the counterparts of a
significant fraction of the still unassociated high-latitude {\it Fermi} $\gamma$-ray 
sources, which make up $\sim 20\%$ of the total at $|b| > 10^{\circ}$. 
These would display a pure elliptical galaxy optical
spectrum and would appear as radio sources typically fainter than currently
known {\it Fermi} steep spectrum blazars (that are also characterized by large LAT error regions) thereby 
failing the association criteria applied by Ref. \refcite{fermi2lac}. 

In conclusion, our scenario is consistent with the complex observational
properties of blazars as we know them from all the surveys carried out so
far in the radio, X-ray, and $\gamma$-ray bands, and solves at the same time a number of
long-standing puzzles. Moreover, it also makes a number of testable predictions. These include, 
for example, the existence of high power -- high \nup~blazars, which would be
very hard to identify because of their featureless optical spectra and,
therefore, lack of redshift. When both \nup~and radio power
are large, in fact, the dilution by the non-thermal continuum becomes extreme and all
optical features are washed away. Four such sources have already been discovered 
\cite{SimplifiedII}. 

\bibliographystyle{ws-procs975x65}
\bibliography{ws-giommi}

\begin{thebibliography}{1}

\bibitem{bla78}
R.~D. {Blandford} and M.~J. {Rees}, {Some comments on radiation mechanisms in
  Lacertids}, in {\em BL Lac Objects\/},  ed. A.~M. {Wolfe} (Pittsburgh, USA,
  1978).

\bibitem{UP95}
C.~M. {Urry} and P.~{Padovani}, {\em PASP} {\bf 107}, p. 803 (1995).

\bibitem{SimplifiedI}
P.~{Giommi}, P.~{Padovani}, G.~{Polenta}, S.~{Turriziani}, V.~{D'Elia} and
  S.~{Piranomonte}, {\em MNRAS} {\bf 420}, 2899 (2012).

\bibitem{SimplifiedII}
P.~{Padovani}, P.~{Giommi} and A.~{Rau}, {\em MNRAS} {\bf 422}, p. L48 (2012).

\bibitem{SimplifiedIII}
P.~{Giommi}, P.~{Padovani} and G.~{Polenta}, {\em MNRAS} {\bf 431}, 1914
  (2013).

\bibitem{ev06}
D.~A. {Evans}, D.~M. {Worrall}, M.~J. {Hardcastle}, R.~P. {Kraft} and
  M.~{Birkinshaw}, {\em ApJ} {\bf 642}, 96 (2006).

\bibitem{fermi2lac}
M.~{Ackermann} and {et al.}, {\em ApJ} {\bf 743}, p. 171 (2011).

\end{thebibliography}

\end{document}